\documentclass[graybox, secnum]{svmult}

\usepackage{mathptmx}       
\usepackage{helvet}         
\usepackage{courier}        
\usepackage{type1cm}        
\usepackage{makeidx}         
\usepackage{graphicx}        
\usepackage{multicol}        
\usepackage[bottom]{footmisc}
\usepackage{hyperref}        
\usepackage{soul}            
\hypersetup{colorlinks=true,urlcolor=blue}
\usepackage[square,numbers]{natbib}
\makeindex             
\usepackage{xcolor}

\usepackage{newtxmath,bm}

\newcommand{\bea}{\begin{eqnarray}}
\newcommand{\eea}{\end{eqnarray}}
\newcommand{\vs}[1]{\vspace{#1 mm}}
\newcommand{\hs}[1]{\hspace{#1 mm}}
\renewcommand{\a}{\alpha}
\renewcommand{\b}{\beta}
\renewcommand{\c}{\gamma}
\renewcommand{\d}{\delta}
\newcommand{\la}{\lambda}
\newcommand{\s}{\sigma}
\newcommand{\G}{\Gamma}
\newcommand{\pa}{\partial}
\newcommand{\Tr}{{\rm Tr}}
\newcommand{\tr}{{\rm tr}}
\newcommand{\Det}{{\rm Det}}
\newcommand{\bg}{\bar g}
\newcommand{\br}{\bar R}
\newcommand{\nn}{\nonumber\\}
\newcommand{\bnabla}{\bar\nabla}
\newcommand{\lich}{{\Delta_L}}

\begin{document}
\title*{One-loop divergences in higher-derivative gravity}
\author{Nobuyoshi Ohta 
}
\institute{Nobuyoshi Ohta \at Department of Physics, National Central University, Zhongli, Taoyuan 320317,
 Taiwan and
Research Institute for Science and Technology, Kindai University, Higashi-Osaka, Osaka 577-8502, Japan,
\email{ohtan@ncu.edu.tw}}
%
%
\maketitle
\abstract{
We give a review of the one-loop divergences in higher derivative gravity theories.
We first make the bilinear expansion in the quantum fluctuation on arbitrary backgrounds,
introduce a higher-derivative gauge fixing and show that higher-derivative gauge fixing must have ghosts
in addition to those naively expected.
We give general formulae for the one-loop divergences in such theories,
and give explicit results for theories with quadratic curvature terms.
In this calculation, we need the heat kernel coefficients for the four-derivative minimal operators and
two-derivative nonminimal vector operators, which are summarized.
We also discuss the beta functions in the renormalization group, and show that the dimensionless
couplings are asymptotically free.
The calculation is also extended to the theories with arbitrary functions of $R$ and $R_{\mu\nu}^2$.
We show that the result is independent of metric parametrization and gauge on shell.
}

\section*{Keywords}

Quantum gravity, Quadratic gravity, Higher-derivative gravity, Perturbation theory, One-loop divergences,
Background method, Heat kernel expansion, Effective action, Renormalization group, Asymptotic freedom

\section{Introduction}

The study of quantum effects in gravity started with the seminal work of 't Hooft and Veltmann~\cite{TV},
in which one-loop divergences were first studied. It was shown that there exist divergences in the quadratic
curvature terms, but these counterterms were shown to be transformed away by a field redefinition.
However it was shown later the Einstein gravity is nonrenormalizable at two loops~\cite{GS}.
It was further shown that gravity theory containing quadratic curvature terms is a renormalizable theory~\cite{Stelle}.
Since this theory contains curvature squares and four derivatives, we refer to such theories as quadratic,
four-derivative or higher-derivative gravity.
Unfortunately the theory is probably nonunitary in perturbation theory, but see \cite{D} for discussions.

The actual calculation of one-loop divergences in the four-derivative quantum gravity involves some complicated
techniques. We use the technique of background field method by separating the quantum fields into backgrounds and
fluctuations, and integrate over the fluctuations. To really do this at one loop, we have to know the bilinear
forms in the fluctuation fields of all relevant terms, such as Weyl curvature square, scalar curvature square
(or other combinations, like Ricci curvature square and Riemann curvature square), and Einstein-Hilbert term.

The next step is to gauge fix the theory. Typically this also involves introduction of the Faddeev-Popov (FP) ghosts
which were first discovered by Feynman~\cite{F}, and introduced in more formal way by DeWitt~\cite{DeWitt},
and later elegantly formulated by Faddeev and Popov~\cite{FP}.
When higher-derivative gauge fixing is chosen, this leads to a complication; in addition to those
naively expected~\cite{JT}, we have to include additional ghost contribution
[denoted as $\Tr\log Y$ in eq.~\eqref{divf}] first noticed in \cite{FT1,FT2},
and derived at the one-loop level in \cite{BC}. Here instead of examining the one-loop result, we give
a more elegant derivation of all the ghosts based on the symmetry principle~\cite{Ohta:2020}.

The general formula for the divergences is then schematically given as
\bea
\Gamma_{\rm div} = \frac12 \Tr \log {\cal H} - \Tr \log \Delta_{gh} - \frac12 \Tr \log Y,
\label{divf}
\eea
where the trace is meant to sum over the spectra of the operators, and the first, second and last terms
are the contributions from the graviton, FP ghosts and additional ghosts, respectively.
To calculate this, it is most convenient to use heat kernel technique, which is explained here in some details.
The calculation involves fourth-order minimal operators in the tensor sector and second-order nonminimal
operators for the ghosts. We give the necessary formulae to evaluate these~\cite{G,GK}.
Then we sum all the contributions to find the one-loop divergences in the quadratic gravity theory.
We mention that there is also the technique called Schwinger-DeWitt method, which is basically the same as
the heat kernel technique~\cite{BV}.
There were some mistakes in the early calculations of the logarithmic divergences~\cite{JT,FT1,FT2},
and the correct results were given in \cite{AB}.

As a simple application of the technique, we also study general higher-derivative theory including
an arbitrary function of Ricci tensor squared and Ricci scalar curvature in a very general parametrization
with arbitrary gauge fixing. We will find the general result for this case, and show that {\it on shell},
it depends on neither the parametrizations nor gauge parameters.


\section{Bilinear expansion of quadratic terms}

We will consider the Euclidean actions of the general form
\bea
S=\int d^d x \sqrt{- g} \Big[ \frac{1}{\kappa^2}(2 \Lambda-R) + \a R^2+\b R_{\mu\nu}^2 + \c R_{\mu\nu\rho\la}^2 \Big] ,
\label{action}
\eea
where $\kappa^2=16\pi G$ is the $d$-dimensional gravitational constant,
$\Lambda$ is the cosmological constant, and $\alpha$, $\beta$, $\gamma$ are the higher-derivative couplings.
Though we consider $d=4$ when we give the results for the divergences, here we give the results
for general dimension $d$.
It is sometimes more convenient to use a different basis for the higher-derivative terms,
namely $R^2$, the square of the Weyl tensor
\bea
C^2 = R_{\mu\nu\a\b}^2-\frac{4}{d-2} R_{\mu\nu}^2+ \frac{2}{(d-1)(d-2)} R^2,
\eea
and the Gauss-Bonnet combination
\bea
E = R_{\mu\nu\a\b}^2- 4 R_{\mu\nu}^2+ R^2,
\eea
which is topological for $d=4$ and vanishes identically for $d=3$.
Conversely we have
\bea
R_{\mu\nu\a\b}^2=\frac{d-2}{d-3}C^2-\frac{1}{d-3}E+\frac{1}{d-1}R^2,~~
R_{\mu\nu}^2=\frac{d-2}{4(d-3)}(C^2-E)+\frac{d}{4(d-1)}R^2.
\eea
Then the action has the alternative form
\bea
S = \int d^d x \sqrt{- g} \Big[ \frac{1}{\kappa^2}( 2 \Lambda-R)
+ \frac{1}{2\la} C^2 - \frac{1}{\rho} E + \frac{1}{\xi} R^2 \Big] ,
\label{action1}
\eea
where
\bea
\lambda = \frac{2(d-3)}{(d-2)(\beta+4\gamma)}, ~~
\rho = \frac{4(d-3)}{(d-2)\beta+4\gamma}, ~~
\xi = \frac{4(d-1)}{4(d-1)\alpha+d\beta+4\gamma}\ .
\eea
or conversely
\bea
\a = -\frac{1}{\rho}+\frac{1}{\xi} +\frac{1}{(d-1)(d-2)\la}, ~~
\b = \frac{4}{\rho} -\frac{2}{(d-2)\la}, ~~
\c = -\frac{1}{\rho} +\frac{1}{2\la}.
\eea
Note that in $d=3$, $C^2$ and $E$ both vanish identically.
The couplings $\lambda$, $\rho$ and $\xi$ have mass dimension $4-d$.
In dimensions higher than three, it is customary to define the dimensionless combinations
\bea
\omega \equiv -\frac{(d-1)\la}{\xi},~~~
\theta \equiv \frac{\la}{\rho}\ .
\eea

We will apply the standard background field method, expanding the metric as
\bea
g_{\mu\nu}= \bg_{\mu\nu} + h_{\mu\nu}\ .
\label{fluc}
\eea
In order to derive the effective action at the one-loop level,
or to calculate the one-loop beta functions,
we need the expansion of the action to second order in $h_{\mu\nu}$.
For this purpose, it is useful to first make the expansion of curvatures in the fluctuations,
which are summarized in Appendix~\ref{expansion}.

Using the formulae given in \ref{expansion}, we find that
the terms proportional to $\alpha$ can be written in the form
\bea
&&
\a h^{\mu\nu} \Big[ \bnabla_\mu \bnabla_\nu \bnabla_\a \bnabla_\b
- \bg_{\mu\nu} \Box \bnabla_\a \bnabla_\b - \bg_{\a\b} \bnabla_\mu \bnabla_\nu \Box
+\bg_{\mu\nu} \bg_{\a\b} \Box^2 -\br \bg_{\nu\b} \bnabla_\a \bnabla_\mu
 \nn &&
- (2 \br_{\mu\nu} - \br \bg_{\mu\nu})\bnabla_\a \bnabla_\b
+ 2 \br_{\mu\nu} \bg_{\a\b}\Box +\frac12 \br (\bg_{\mu\a}\bg_{\nu\b}
- \bg_{\mu\nu}\bg_{\a\b})\Box -\br \br_{\a\b}\bg_{\mu\nu} \nn
&&
+ 2\br \br_{\mu\a} \bg_{\b\nu} + \br_{\mu\nu} \br_{\a\b}
-\frac{1}{4}J_{\mu\nu\alpha\beta}\br^2
\Big] h^{\a\b}\ .
\label{r1}
\eea
Here and in what follows, $\Box \equiv \bnabla_\mu \bnabla^\mu$ and a bar indicates that the quantity is
evaluated on the background;
the indices are raised, lowered and contracted by the background metric $\bg$,
and the covariant derivative $\bnabla$ is constructed with the background metric.
The tensor $J$ is defined by
\bea
J_{\mu\nu\a\b}=\d_{\mu\nu,\a\b}-\frac12 \bg_{\mu\nu} \bg_{\a\b}
\label{defJ}
\eea
where
\bea
\d_{\mu\nu,\a\b} = \frac12 (\bg_{\mu\a} \bg_{\nu\b} + \bg_{\mu\b} \bg_{\nu\a})
\equiv \hat 1,
\label{id}
\eea
is the identity in the space of symmetric tensors.

The $\b$ terms can be written in the form
\bea
\b h^{\mu\nu}\hs{-2}&& \Big[ \frac12 \bnabla_\mu \bnabla_\nu \bnabla_\a \bnabla_\b
- \frac12 \bg_{\mu\nu} \bnabla_\a \Box \bnabla_\b
-\frac12 \bg_{\nu\b} \bnabla_\mu \Box \bnabla_\a +\frac14( \bg_{\mu\a} \bg_{\nu\b}
+\bg_{\mu\nu} \bg_{\a\b})\Box^2
\nn &&
+\frac12 \br_{\nu\b} \bnabla_\a \bnabla_\mu
-2 \br_\mu^\rho \bg_{\nu\b} \bnabla_\rho\bnabla_\a
+ \frac32 \bg_{\a\b}\br_{\rho\mu} \bnabla^\rho \bnabla_\nu
+ \br_{\mu\a\nu\b} \Box
\nn &&
+\frac14 (2 \bg_{\mu\a} \bg_{\nu\b}
- \bg_{\mu\nu} \bg_{\a\b}) \br^{\rho\la}\bnabla_\rho \bnabla_\la
-\frac32 \br^\rho_\b \br_{\rho\mu\nu\a}
+ \br_{\mu\rho\la\nu} \br_\a{}^{\rho\la}{}_\b
\nn &&
+\frac12 \bg_{\nu\b} \br_{\mu\rho}\br^{\rho}_\a
- \bg_{\a\b} \br_{\mu\rho} \br^\rho_\nu
-\frac{1}{4}J_{\mu\nu\alpha\beta}\br_{\rho\sigma}^2
\Big] h^{\a\b}\ ,
\label{r2}
\eea
and the terms proportional to $\c$ are
\bea
\c h^{\mu\nu} \hs{-2}&& \Big[ \bnabla_\mu \bnabla_\nu \bnabla_\a \bnabla_\b
+ \bg_{\mu\a} \bg_{\nu\b} \Box^2
-2 \bg_{\nu\b}\bnabla_\mu \Box \bnabla_\a
- 2 \bg_{\nu\b} \br_{\mu\rho\la\a} \bnabla^\rho \bnabla^\la
+ 3 \br_{\mu\a\nu\b}\Box \nn
&&
- 4 \br_{\mu \a\rho\nu} \bnabla^\rho \bnabla_\b
- 4 \bg_{\nu\b} \br_{\rho\mu} \bnabla^\rho \bnabla_\a
-2\bg_{\nu\b}\br_{\mu\a}\Box
+ \bg_{\mu\a} \bg_{\nu\b} \br_{\rho\la} \bnabla^\rho \bnabla^\la \nn
&& -2\bg_{\mu\nu} \br_{\rho\a\la\b}\bnabla^\la\bnabla^\rho
+ 4 \br_{\mu\a} \bnabla_\nu \bnabla_\b
+ 2 \bg_{\nu\b} \br_{\mu\la\rho\s} \br_\a{}^{\la\rho\s}
-2 \bg_{\nu\b} \br^{\rho\la} \br_{\mu\rho\a\la}
\nn
&&
- \br_{\mu\la\rho\b} \br_\nu{}^{\rho\la}{}_\a
+ 3 \br_{\mu\a}{}^{\rho\la} \br_{\nu\rho\b\la}
-3 \br_{\mu\la\nu\rho} \br_\a{}^{\rho\la}{}_\b
-3 \br^\rho_\a \br_{\mu\b\nu\rho}
\nn
&&
+2 \br_{\mu\a} \br_{\nu\b}
- \bg_{\mu\nu} \br_\a{}^{\rho\la\s} \br_{\b\rho\la\s}
-\frac{1}{4}J_{\mu\nu\alpha\beta}\br_{\rho\sigma\lambda\tau}^2
\Big] h^{\a\b}.
\label{r3}
\eea
It can be checked that when arranged in the Gauss-Bonnet combination
($\gamma=\alpha$, $\beta=-4\alpha$)
and the background metric is maximally symmetric, one obtains a total derivative.
This gives a nontrivial check of the results.

\section{Gauge fixing and ghosts}

The first attempt at calculating one-loop divergences was made in \cite{JT} with higher-order gauge fixing,
but it was pointed out that it did not correctly incorporate the FP ghosts. The problem is that an additional
ghost contribution was missing, which was first considered in \cite{FT1,FT2}. But it was not clear why we need
such additional ghosts, and later clarified in \cite{BC}; this showed it must be there by carefully examining
the one-loop amplitude in the path integral formulation.
Here we explain more elegant way of introducing the correct FP ghosts including the additional ghosts for
higher-derivative gauge fixing~\cite{Ohta:2020}, which is valid not only for one loop but also for all loops
because it is based on the exact BRST (Becchi-Rouet-Stora-Tyutin) symmetry of the system.

The BRST transformation for the fields is found to be
\bea
\d_B g_{\mu\nu} &=& -\d \la [ g_{\rho\nu}\nabla_\mu c^\rho + g_{\rho\mu}\nabla_\nu c^\rho]
\equiv -\d\la{\cal D}_{\mu\nu,\rho}c^\rho, \nn
\d_B c^\mu &=& -\d\la c^\rho \nabla_\rho c^\mu,~~~
\d_B \bar c_\mu = i \d\la\, B_\mu, ~~~
\d_B B_\mu = 0,
\label{brst}
\eea
which is nilpotent. Here $c^\mu, \bar c_\mu$ and $B_\mu$ are the FP ghost,
anti-ghost and an auxiliary field, respectively, and $\d\la$ is an anticommuting parameter.
The gauge fixing term and the FP ghost terms are concisely written as
\bea
{\cal L}_{GF+FP}/\sqrt{-\bg} &=& -i \d_B [\bar c_\mu Y^{\mu\nu}
(\chi_\nu-\frac{a}{2} B_\nu)]/\d\la \nn
\hs{-2} &=& B_\mu Y^{\mu\nu} \chi_\nu
- i \bar c_\mu Y^{\mu\nu} ( \nabla^\la{\cal D}_{\la\nu,\rho}+ b \nabla_\nu{\cal D}_{\la,\rho}^\la) c^\rho
-\frac{a}{2} B_\mu Y^{\mu\nu} B_\nu \nn
&=& \frac{1}{2a} \chi_\mu Y^{\mu\nu} \chi_\nu - \frac{a}{2}{\tilde B}_\mu Y^{\mu\nu}{\tilde B}_\nu
+ i \bar c_\mu Y^{\mu\nu} \Delta_{gh, \nu \rho}c^\rho,
\label{gfgh}
\eea
where
\vs{-3}
\bea
\chi_\mu \equiv \nabla^\la h_{\la\mu} + b \nabla_\mu h,\hs{5}
\tilde B_{\mu} \equiv B_\mu -\frac{1}{a}\chi_\mu,
\eea
are the gauge fixing function and a field imposing the gauge condition, respectively,
$Y^{\mu\nu}$ is a derivative operator for higher-derivative gauge fixing
\bea
Y_{\mu\nu} \equiv -\bg_{\mu\nu} \Box - c \nabla_\mu \nabla_\nu + d \nabla_\nu \nabla_\mu,\hs{3}
\Delta_{gh,\mu\nu} \equiv -\bg_{\mu\nu}\Box -(1+2b)\nabla_\mu\nabla_\nu - R_{\mu\nu},~~
\label{fpghost}
\eea
$\Delta_{gh,\mu\nu}$ is the ghost kinetic term,
and $a$, $b$, $c$ and $d$ are gauge parameters. For one-loop calculation, we can replace $\nabla$ by $\bnabla$.
We see from \eqref{gfgh} that we get the determinant factor $(\Det~Y^{\mu\nu})^{1/2}$ after we perform
the path integral over $\tilde B_\mu$ and FP ghosts $\bar c_\mu$ and $c^\rho$.
(The factor $Y^{\mu\nu}$ in the first term combines into the graviton contribution.)
In the standard way of Faddeev and Popov, the factor of $Y$ can be easily missed~\cite{JT},
but in this formulation we see why there must be this factor. Note also that the field $B_\mu$
is an auxiliary field for low-derivative gauge fixing, but here it becomes dynamical due to the higher-derivative
gauge fixing with the factor $Y^{\mu\nu}$.

We choose the gauge parameters such that the nonminimal four derivative terms
$\bnabla_\mu \bnabla_\nu \bnabla_\a \bnabla_\b$, $\bg_{\mu\nu} \Box\bnabla_\a \bnabla_\b$
and $\bg_{\nu\b} \bnabla_\mu \Box \bnabla_\a$ cancel. This leads to the choice~\cite{BS2,OP1}
\bea
a= \frac{1}{\b+4\c},~~~
b= \frac{4\a+\b}{4(\c-\a)}, ~~~
c-d = \frac{2(\c-\a)}{\b+4\c}-1.
\eea
In order to simplify the gauge-fixing term, we will further choose $d=1$.
Then, the quadratic terms in the action can be written in the form
$h^{\mu\nu} {\cal K}_{\mu\nu,\a\b} h^{\a\b}$,
where
\bea
{\cal K}=K \Box^2 + D_{\rho\la} \bnabla^\rho \bnabla^\la + W.
\label{quartic operator}
\eea
The explicit forms of the coefficients are
\bea
(K)_{\mu\nu,\a\b} = \frac{\b+4\c}{4} \Big(\bg_{\mu\a}\bg_{\nu\b}
+\frac{4\a+\b}{4(\c-\a)}\bg_{\mu\nu}\bg_{\a\b} \Big),
\eea
\vs{-5}
\bea
&& \hs{-5} (D_{\rho\la})_{\mu\nu,\a\b} \nn
&=&
-2\c \bg_{\nu\b}\br_{\a\rho\la\mu}
+4\c \bg_{\rho\nu} \br_{\la\a\mu\b}
+(\b + 3\c) \bg_{\rho\la} \br_{\mu\a\nu\b}
-(2\b+4\c) \bg_{\a\rho}\bg_{\nu\b} \br_{\mu\la} \nn
&& \hs{-2} -2\c \bg_{\nu\b} \br_{\mu\a} \bg_{\rho\la}
+ \b \bg_{\mu\nu}\bg_{\b\rho} \br_{\a\la}
-2\a \bg_{\a\rho} \bg_{\b\la} \br_{\mu\nu}
+2\a \bg_{\mu\nu} \bg_{\rho\la} \br_{\a\b}
+2 \c \bg_{\mu\nu} \br_{\a\rho\la\b} \nn
&&
+ \Big(\frac{\a}{2}\br -\frac{1}{4\kappa^2} \Big) (\bg_{\mu\a} \bg_{\nu\b} \bg_{\rho\la}
- \bg_{\mu\nu} \bg_{\a\b} \bg_{\rho\la}
- 2 \bg_{\nu\b} \bg_{\mu\rho} \bg_{\a\la}
+ 2 \bg_{\mu\nu} \bg_{\a\rho} \bg_{\b\la}) \nn
&& \hs{-2}  +2\c \bg_{\nu\rho} \bg_{\b\la} \br_{\mu\a}
+\Big(\frac{\b}{2}+\c \Big) \bg_{\mu\a} \bg_{\nu\b} \br_{\rho\la}
-\frac{\b}{4} \bg_{\mu\nu} \bg_{\a\b} \br_{\rho\la},
\label{der}
\eea
\bea
&& (W)_{\mu\nu,\a\b} \nn
&&=
\frac32 \c \bg_{\nu\b} \br_\mu{}^{\rho\la\s} \br_{\a\rho\la\s} 
+ 4\c \br_{\rho\a\mu\la} \br_{\nu\b}{}^{\rho\la}
- \c \br_{\rho\a\mu\la} \br^\rho{}_{\nu\b}{}^\la
+(\b+5\c) \br_{\rho\mu\la\nu} \br^\rho{}_\a{}^\la{}_\b \nn
&& \hs{-2} + 6\c \br^\rho_\mu \br_{\rho\a\b\nu}
+ \Big(\frac{\b}{2}+\c \Big) \br_{\mu\a} \br_{\nu\b}
+\Big(\a\br - \frac{1}{2\kappa^2}\Big)\Big(\frac12\br_{\mu\a\nu\b}+
\frac32\bg_{\b\nu}\br_{\mu\a} -  \bg_{\mu\nu}\br_{\a\b}\Big) \nn
&& \hs{-2} + \a \br_{\mu\nu} \br_{\a\b}
+ \frac{1}{8} \Big(\a \br^2 +\b \br_{\rho\la}^2 +\c \br_{\rho\la\s\tau}^2
-\frac{1}{\kappa^2}(\br-2\Lambda) \Big)
(\bg_{\mu\nu} \bg_{\a\b}-2\bg_{\mu\a} \bg_{\nu\b} ) \nn
&& \hs{-2} +\Big(\frac{5}{2}\b+4\c \Big) \bg_{\nu\b} \br_{\mu\rho} \br^\rho_\a
- \c \bg_{\mu\nu} \br_\a{}^{\rho\la\s} \br_{\b\rho\la\s}
- \b \bg_{\a\b} \br_{\mu\rho} \br^\rho_\nu
- (\b+4\c) \bg_{\nu\b} \br^{\rho\la} \br_{\mu\rho\a\la},\nn
\label{curm}
\eea
where we have dropped terms with two derivatives acting on a background curvature,
and performed the symmetrizations $\mu\leftrightarrow\nu$, $\alpha\leftrightarrow\beta$ and
$(\mu,\nu)\leftrightarrow (\alpha,\beta)$.

In order to use the heat kernel formula, we have to put this operator into the form
\bea
{\cal H}=K^{-1}{\cal K}
=\Box^2 + V_{\rho\la} \bnabla^\rho \bnabla^\la +U\ ,
\label{hami}
\eea
where
\bea
(K^{-1})_{\mu\nu}{}^{\a\b}
=\frac{4}{\b+4\c} (\d_{\mu\nu}{}^{\a\b} -\Omega \bg_{\mu\nu} \bg^{\a\b}),
\eea
with
\bea
\Omega = \frac{4\a+\b}\Sigma, ~~~
\Sigma \equiv 4(\c-\a)+d(4\a+\b).
\label{sigma}
\eea
The form of the coefficients $V_{\rho\lambda}$ and $U$ is complicated.
First $V$ is given by
\newcommand{\bk}{{\bf k}}
\bea
V^{\rho\la} = \frac{4}{\b+4\c} \sum_{i=1}^{20} b_i \bk_i,
\eea
where
\bea
&& \bk_1= \bg_{\nu\b} \bg^{\rho\la} \br_{\mu\a},~~~
\bk_2= \d_{\mu\nu,\a\b} \bg^{\rho\la},~~~
\bk_3= \bg^{\rho\la} \br_{\mu\a\nu\b},~~~
\bk_4= \d_{\nu\b}{}^{\rho\la} \br_{\mu\a}, \nn
&& \bk_5= \d_{\nu\b}{}^{\rho\la} \bg_{\mu\a},~~~~
\bk_6= \d_{\mu\nu,\a\b}\br^{\rho\la},~~~
\bk_7= \frac12 (\d_\nu^{(\rho} \br^{\la)}{}_{\a\b\mu} + \d_\b^{(\rho} \br^{\la)}{}_{\mu\nu\a}),\nn
&& \bk_8= \bg_{\nu\b}\d^{(\rho}_{(\mu} \br^{\la)}_{\a)}, ~~~
\bk_9 = \bg_{\nu\b} \br_{(\a}{}^{\rho\la}{}_{\mu)},~~~
\bk_{10}= \frac12 (\d_{\a\b}{}^{\rho\la}\br_{\mu\nu} + \d_{\mu\nu}{}^{\rho\la} \br_{\a\b}),\nn
&& \bk_{11} = \bg_{\mu\nu} \br_{\a}{}^{\rho\la}{}_{\b},~~~
\bk_{12} = \bg_{\a\b} \br_{\mu}{}^{\rho\la}{}_{\nu},~~~
\bk_{13} = \bg_{\mu\nu} \bg^{\rho\la} \br_{\a\b},~~~
\bk_{14} = \bg_{\a\b} \bg^{\rho\la} \br_{\mu\nu}, \nn
&& \bk_{15} = \bg_{\mu\nu} \d_\a^\la \br_\b^\rho,~~~~
\bk_{16} = \bg_{\a\b} \d_\mu^\la \br_\nu^\rho,~~~
\bk_{17}= \bg_{\mu\nu} \d_{\a\b}{}^{\rho\la},~~~
\bk_{18}= \bg_{\a\b} \d_{\mu\nu}{}^{\rho\la}, \nn
&& \bk_{19} = \bg_{\mu\nu} \bg_{\a\b} \bg^{\rho\la}, ~~~
\bk_{20} = \bg_{\mu\nu} \bg_{\a\b} \br^{\rho\la},~~~
\eea
and
\bea
&& b_1= -2\c, ~~~
b_2=\frac{\a}{2} \br -\frac{1}{4\kappa^2}, ~~~
b_3 = \b+3\c, ~~~
b_4 = 2\c, ~~~
b_5 = \frac{1}{2\kappa^2} -\a \br,\nn
&& b_6 = \frac{\b}{2}+\c,~~~
b_7 = -4\c, ~~~
b_8 = -2\b-4\c, ~~~
b_9 = -2\c, ~~~
b_{10} = -2\a, \nn
&& b_{11} = 4\c \Omega_3,~~~
b_{12} = \c, ~~~
b_{13} = -\b \Omega_3, ~~~
b_{14} = \a, ~~~
b_{15} = 2\b \Omega_3,~~~
b_{16} = \frac{\b}{2}, \nn
&& b_{17} = 2\a \Omega_3 \br - \frac{\Omega_1-2\Omega}{2\kappa^2}, ~~~
b_{18} = \frac{\a}{2}\br-\frac{1}{4\kappa^2}, ~~~
b_{19}=-b_{17}, ~~
b_{20}=-\b \Omega_3.~~
\label{v}
\eea
where
\bea
\Omega_1 = \frac{10\a+3\b+2\c}{\Sigma}, ~~~
\Omega_3 = \frac{3\a+\b+\c}{\Sigma}, ~~~
\eea
with $\Sigma$ given in \eqref{sigma}.
Next,
\bea
&& (U)_{\mu\nu,\a\b} \nn
&&= \frac{4}{\b+4\c} \Bigg[
\frac32 \c \bg_{\nu\b}\br_\mu{}^{\rho\la\s} \br_{\a\rho\la\s}
-\c \br^\la{}_{\a\mu}{}^\rho \br_{\la\nu\b\rho}
+4\c \br_{\rho\a\mu\la} \br_{\nu\b}{}^{\rho\la} \nn
&& -3\c(\br_\mu^\s \br_{\s\a\nu\b} + \br^\s_\a \br_{\s\mu\b\nu})
+\Big(\frac{\b}{2}+\c \Big) \br_{\mu\a}\br_{\nu\b}
-\frac{\c}{2}\bg_{\a\b} \br_{\mu\rho\la\s} \br_\nu{}^{\rho\la\s} \nn
&& +\frac14 S^2 (\Omega_1 \bg_{\mu\nu} \bg_{\a\b} -\bg_{\mu\a} \bg_{\nu\b} )
+\Big(\frac{\a}{2}\br -\frac{1}{4\kappa^2} \Big)
 (\br_{\mu\a\nu\b}+ 3\bg_{\nu\b} \br_{\mu\a} -\bg_{\a\b} \br_{\mu\nu})\nn
&& +\Big(\frac{5}{2}\b+4\c \Big) \bg_{\nu\b} \br_{\mu\s} \br^{\s}_\a
+(\b+5\c) \br_{\rho\mu\la\nu} \br^\rho{}_\a{}^\la{}_\b
- \frac{\b}{2} \bg_{\a\b} \br_{\mu\s}\br^\s_\nu \nn
&& - \c \Omega_1 \bg_{\mu\nu} \br_{\a\rho\la\s} \br_\b{}^{\rho\la\s}
- \b \Omega_1 
 \bg_{\mu\nu} \br_{\a\s}\br^\s_\b
+ \a \br_{\mu\nu} \br_{\a\b}
+\Big( \frac{1}{\kappa^2} \Omega_3 -\a \Omega_1 \br \Big) \bg_{\mu\nu} \br_{\a\b} \nn
&& +\frac{1}{4\kappa^2}(\br-4\Lambda)\Omega \bg_{\mu\nu} \bg_{\a\b}
-(\b+4\c) \bg_{\nu\b} \br^{\rho\la}\br_{\mu\rho\a\la} \Bigg],
\eea
where we have defined
\bea
S^2 = \a \br^2+\b \br_{\mu\nu}^2 + \c \br_{\mu\nu\rho\la}^2
- \frac{1}{\kappa^2}(\br-2 \Lambda),
\eea
These results are given in Refs.~\cite{BS2,OP1}.

\section{General formula for one-loop divergences}

Correcting the errors in \cite{JT}, the results for one-loop divergences were given in \cite{FT1,FT2},
but they still contained errors, and this was corrected in \cite{AB}.
Here we give the correct results using heat kernel expansion.

The partition function for the one-loop is obtained from the quadratic terms of the action as
\bea
Z=\Det(\Delta)^{-1/2},
\eea
for a real scalar field. and the effective action is given by
\bea
\G=-\log Z = \frac12 \Tr\log \Delta.
\label{eff}
\eea
If the fluctuations are anticommuting fields like the FP ghosts, the sign should be opposite.
If the fluctuations are complex fields (or independent two hermitian fields), the front factor should be 1.

Suppose we know the eigenvalues and eigenfunctions of the operator $\Delta$:
\bea
\Delta\phi_n=\la_n \phi_n.
\eea
Then we can evaluate \eqref{eff} as
\bea
\frac12 \Tr\log \Delta=\frac12 \sum_n \log \la_n.
\eea
We define the zeta function for $\Delta$:
\bea
\zeta_\Delta(s)=\sum_{n=1}^\infty \la_n^{-s} =\frac{1}{\G(s)}\int_0^\infty dt\, t^{s-1} \Tr (e^{-t\Delta}),
\eea
and obtain
\bea
\frac12 \Tr\log \Delta=\left. -\frac12 \frac{d}{ds}\zeta_\Delta(s)\right|_{s=0}
=-\frac12 \frac{d}{ds}\left[ \frac{1}{\G(s)}\int_0^\infty dt \, t^{s-1} \Tr (e^{-t\Delta}) \right]_{s=0}.
\eea

For a differential operator of order $p$ in $d$ dimensions, the heat kernel $e^{-t\Delta}$ has the expansion
\bea
\Tr (e^{-t\Delta}) = \int \frac{d^d x}{(4\pi)^{d/2}} \sqrt{g} \sum_{n=0}^\infty b_{2n} (\Delta)\, t^{(2n-d)/p}.
\label{heatkernel}
\eea

Typically we have the operators of $p=4$ and $p=2$. The formulae are given separately.
The divergent part of the effective action for $p=4$ operator is evaluated as
\bea
\G^{(4)} &=& -\frac12 \int \frac{d^d x}{(4\pi)^{d/2}} \sqrt{g} \int_{1/\Lambda_{\rm UV}^4}^{1/\mu^4} dt
\left[ t^{-\frac{d}{4}-1} b_0 + t^{-\frac{d-2}{4}-1} b_2 +\cdots + t^{-1} b_{d} + \cdots \right] \nn
&=& -\frac12 \int \frac{d^d x}{(4\pi)^{d/2}} \sqrt{g}
\left[ \frac{\Lambda_{\rm UV}^d}{d/4} b_0 + \frac{\Lambda_{\rm UV}^{d-2}}{(d-2)/4} b_2 +\cdots
+ \log\frac{\Lambda_{\rm UV}^4}{\mu^4} b_d
+ \mbox{ finite } \right],
\qquad
\label{div4}
\eea
where the heat kernel coefficients are given in the subsection~\ref{p=4minimal}.
Here $\Lambda_{\rm UV}$ is an ultraviolet cutoff which should be distinguished from the cosmological constant $\Lambda$.

The divergent part of the effective action for $p=2$ operator is given by
\bea
\G^{(2)} &=& -\frac12 \int \frac{d^d x}{(4\pi)^{d/2}} \sqrt{g} \int_{1/\Lambda_{\rm UV}^2}^{1/\mu^2} dt
\left[ t^{-\frac{d}{2}-1} b_0 + t^{-\frac{d}{2}} b_2 +\cdots + t^{-1} b_d + \cdots \right] \nn
&=& -\frac12 \int \frac{d^d x}{(4\pi)^{d/2}} \sqrt{g}
\left[ \frac{\Lambda_{\rm UV}^d}{d/2} b_0 + \frac{\Lambda_{\rm UV}^{d-2}}{\frac{d}{2}-1} b_2 +\cdots
+ \log\frac{\Lambda_{\rm UV}^2}{\mu^2} b_d + \mbox{ finite } \right] .
\qquad
\label{div2}
\eea
These are relevant for the contributions from the ghost and the operator $Y$.

In our present case, the graviton contribution~\eqref{hami} is $p=4$ and the ghost contributions are $p=2$.
So the one-loop part of our effective action is
\bea
\G^{\rm 1-loop} &=& \frac12 \Tr\log{\cal H}- \Tr\log{\Delta_{gh}} -\frac12 \Tr\log{Y}
\nn
&=& - \int\frac{d^4 x}{2(4\pi)^2}\sqrt{g} \Big[
\Lambda_{\rm UV}^4 \Big( b_0({\cal H}) -b_0(\Delta_{gh})-\frac12 b_0(Y) \Big) \nn
&& \hs{10} + \Lambda_{\rm UV}^2 \Big( 2b_2({\cal H}) -2 b_2(\Delta_{gh})-b_2(Y) \Big) \nn
&& \hs{10} + \log\frac{\Lambda_{\rm UV}^2}{\mu^2}\Big( 2b_4({\cal H})- 2 b_4(\Delta_{gh}) -b_4(Y) \Big) \Big].
\eea

Now we start the evaluation of these contributions.

\subsection{Contributions from $p=4$ minimal operator}
\label{p=4minimal}

First we have to evaluate the contribution from \eqref{hami} which is $p=4$ minimal operator.
The coefficients in the heat kernel expansion for spin 2 are~\cite{G}:
\bea
b_0({\cal H}) &=&  \frac{\Gamma(d/4)}{2\Gamma(d/2)}\frac{d(d+1)}{2} \;, \\
b_2({\cal H}) &=&  \frac{\Gamma((d-2)/4)}{2\Gamma((d-2)/2)}
\,\mathrm{tr}\left[\frac{\br}{6}+\frac{1}{2d}V^\mu_\mu\right],\\
b_4 ({\cal H}) &=&  \frac{\Gamma(d/4)}{2\Gamma((d-2)/2)}
\mathrm{tr}\Bigl[ \frac{\hat 1}{90} \br_{\rho\la\s\tau}^2 -\frac{\hat 1}{90} \br_{\rho\la}^2
+\frac{\hat 1}{36}\br^2 +\frac{1}{6} {\Omega}_{\rho\la}{\Omega}^{\rho\la} - \frac{2}{d-2} U \nn
&& - \frac{1}{6(d-2)} (2 \br_{\rho\la}V^{\rho\la} - \br V^\rho{}_\rho)
 + \frac{1}{4(d^2-4)} (V^\rho{}_\rho V^\la{}_\la + 2 V_{\rho\la} V^{\rho\la})
\nn && +\frac{1}{15}\Box \br+\frac{d+4}{6(d^2-4)}\Box V^\rho{}_\rho
-\frac{2(d+1)}{3(d^2-4)}\bnabla^\rho\bnabla^\la V_{\rho\la} \Bigr],
\label{grav}
\eea
where $\hat 1$ is the identity defined in \eqref{id} and
${\Omega}_{\rho\la}$ is the commutator of the covariant derivatives acting
on the tensor $h^{\a\b}$: ${\Omega}_{\rho\la} = [\bnabla_\rho, \bnabla_\la ]$.
The traces should be taken over the space of symmetric tensors with the identity~\eqref{id}.
These give
$\tr(\hat 1)=\frac{d(d+1)}{2},\, \tr({\Omega}_{\rho\la}{\Omega}^{\rho\la})=-(d+2) \br_{\mu\nu\a\b}^2$.

We need more explicit formulae for the traces. Because these are very complicated for general
dimensions, we give the results for $d=4$ and omit total derivative terms. We find
\bea
\tr\, U = \d^{\mu\nu,\a\b} U_{\mu\nu,\a\b}
= A_1 \br_{\mu\nu\rho\la}^2 + A_2 \br_{\mu\nu}^2
+ A_3 \br^2-A_4 \frac{\br}{\kappa^2}-A_5 \frac{\Lambda}{\kappa^2},
\eea
where
\bea
A_1=3,~~
A_2=\frac{8}{3}+\frac{4\la}{\xi},~~
A_3=\frac{1}{3}+\frac{2\la}{\xi}, ~~
A_4=3\la,~~
A_5=\frac{2}{9}(84\la-\xi).
\label{coea}
\eea
and
\bea
\tr\, (V^\rho_\rho \br)
= B_1 \br^2 - B_2 \frac{\br}{\kappa^2},
\eea
where
\bea
B_1=-\frac{68}{3}+\frac{32\la}{\xi},~~
B_2=-20\la+\frac{2\xi}{3}.
\label{coeb}
\eea
Next
\bea
\tr\, (V^{\rho\la} \br_{\rho\la})
= C_1 \br_{\mu\nu}^2 + C_2 \br^2-C_3 \frac{\br}{\kappa^2},
\eea
where
\bea
C_1=\frac{8}{3}-\frac{8\la}{\xi},~~
C_2=-\frac{19}{3}+\frac{10\la}{\xi},~~
C_3=-5\la+\frac{\xi}{6}.
\label{coec}
\eea
Finally
\bea
\frac{1}{48} \tr (V^\rho_\rho V^\la_\la)
+\frac{1}{24} \tr (V_{\rho_\la} V^{\rho\la})
=D_1 \br_{\mu\nu\rho\la}^2 + D_2 \br_{\mu\nu}^2+D_3 \br^2-D_4 \frac{\br}{\kappa^2}
+D_5 \frac{1}{\kappa^4},~~~~
\eea
where
\bea
&& D_1=6,~~
D_2=\frac{2(18\la^2+6\la\xi+113\xi^2)}{27\xi^2},~~
D_3=\frac{576\la^2-240 \la\xi -47 \xi^2}{54\xi^2},\nn
&& D_4=\frac{-180 \la^2 + 30\la\xi + \xi^2}{18\xi},~~
D_5=\frac{180\la^2+\xi^2}{72}.
\label{coed}
\eea

\subsection{Contribution from the $p=2$ nonminimal vector operator}
\label{p=2nonminimalvector}

To find the contribution from the ghost operator, we need the contribution from $p=2$ nonminimal vector operator.
The general form of the $p=2$ nonminimal operator is
\bea
\Delta= - \bg^{\mu\nu} \Box+a \bnabla^\mu \bnabla^\nu +X^{\mu\nu},
\label{p=2}
\eea

The coefficients in the heat kernel expansion~\eqref{heatkernel} have been calculated in \cite{GK}.
For the general case, we have
\bea
b_0 = (1-a)^{-d/2}+d-1, \hs{80}
\eea
\vs{-5}
\bea
b_2 =  \left(\frac{d^2-d-6}{6d}+(1-a)^{-d/2}\frac{6+(1-a)d}{6d} \right) \br
 +\frac{1-d-(1-a)^{-d/2}}{d} X,\hs{17}
\label{2b2}
\eea
\bea
\hs{-4}
b_4 &=& \frac{d-16+(1-a)^{2-d/2}}{180} \br_{\mu\nu\rho\la}^2
\nn &&
+\frac{1}{180 ad(d^2-4)}\Big[-360d-(d^4-d^3+116d^2-296d-360)a
 +(1-a)^{-d/2}[-360a
\nn &&
\hs{10} +4(90-74a+28a^2+a^3)d -60a(1-a)d^2-(1-a)^2ad^3] \Big] \br_{\mu\nu}^2
\nn &&
+\frac{1}{72 ad(d^2-4)}\Big[72(2-a) +(16d-16d^2-d^3+d^4)a
\nn &&
+(1-a)^{-d/2}[-72(2-a)-4a(a^2+4a-14)d
+12a(1-a)d^2+a(1-a)^2 d^3] \Big] \br^2
\nn &&
+\frac{1}{6ad(d^2-4)}\Big[ \Big\{-24+12a+ad^2-ad^3+(1-a)^{-d/2}[24-12a
\nn &&
+2a(a-4)d-a(1-a)d^2] \Big\} \br X
+2 \Big\{-12a+4(3-2a)d+5ad^2
\nn &&
+(1-a)^{-d/2}[12a-2(6-4a+a^2)d+a(1-a)d^2] \Big\} \br_{\mu\nu} X^{\mu\nu}
\nn &&
+3 \Big\{4-2a+ad+(1-a)^{-d/2}(-4+2a+ad) \Big\} X^2 +3\Big\{4a-2(2+a)d-2ad^2
\nn &&
+ad^3+(1-a)^{-d/2}[-4a+2(2-a)d]\Big\} X_{\mu\nu}X^{\mu\nu} \Big]  +\mbox{ (total derivative terms)},
\eea
where $X=X_\mu^\mu$.
\\

\noindent
\underline{\bf Contribution from the the ghost operator $\Delta_{gh}$}
\\

For the ghost operator $\Delta_{gh}$ in \eqref{fpghost},
we see that we have $a=-1-\frac{4\a+\b}{2(\c-\a)}\equiv \s_g$ and $X_{\mu\nu}=-\br_{\mu\nu}$ in \eqref{p=2}.
The above formulae give, for $d=4$,
\bea
b_0(\Delta_{gh}) &=& (1-\s_g)^{-2}+3, \\
b_2(\Delta_{gh}) &=& \frac{6\s_g^2 -13\s_g+10}{6(1-\s_g)^2} \br, \\
b_4(\Delta_{gh}) &=& \frac{-11}{180} \br_{\mu\nu\rho\la}^2
-\frac{2\sigma_g^2+26\sigma_g -43}{90(1-\sigma_g )^{2}}\br_{\mu\nu}^2
+\frac{5\sigma_g^2 - 10\sigma_g +8}{36 (1-\sigma_g )^{2}} \br^2.
\eea

\noindent
\underline{\bf Contribution from the the additional ghost operator $Y$}
\\

For the operator $Y$ in~\eqref{fpghost}, we first note that
$[\bnabla_\mu,\bnabla_\nu]\chi^\nu = 
-\br_{\mu\nu} \chi^\nu$. Using this relation, we find that
we have $a=1+2\frac{\a-\c}{\b+4\c}\equiv \s_Y$ and
$X_{\mu\nu}=\br_{\mu\nu}$ in \eqref{p=2}, and the above formulae give, for $d=4$,
\bea
b_0(Y) &=& (1-\s_Y)^{-2}+3, \\
b_2(Y) &=& \frac{3\s_Y-2}{6(1-\s_Y)} \br, \\
b_4(Y) &=& \frac{-11}{180} \br_{\mu\nu\rho\la}^2
+\frac{43}{90} \br_{\mu\nu}^2 -\frac{1}{9} \br^2.
\eea

\section{One-loop divergences and asymptotic freedom}

We are now ready to calculate the divergences in the quadratic curvature theory on the general backgrounds.
For this purpose, we need to know the heat kernel coefficients.
In this section, we restrict the dimension of our spacetime to four.

Putting the results for ${\cal H}$ into eq.~\eqref{grav}, we get
\bea
b_0({\cal H}) &=& 5,
\\
b_2({\cal H}) &=& \frac{\sqrt{\pi}}{2} \left[\frac{5\br}{3}+\frac{1}{8}\Big(B_1 \br-B_2\frac{1}{\kappa^2}\Big)\right],
\label{qb2}
\\
b_4({\cal H}) &=& \frac{1}{2} \Biggl[
\br_{\mu\nu\rho\la}^2\left(-\frac{8}{9}- A_1+D_1\right)
-\br_{\mu\nu}^2\left(\frac{1}{9}+A_2+\frac{1}{6}C_1-D_2\right)
\nn
&&
+\br^2 \left(\frac{5}{18}-A_3+\frac{1}{12}B_1-\frac{1}{6}C_2+D_3\right)
+\frac{1}{\kappa^2}\br\left(A_4-\frac{1}{12}B_2+\frac{1}{6}C_3-D_4\right)
\nn
&&
+\frac{1}{\kappa^2}\left(\Lambda\,A_5+\frac{1}{\kappa^2}D_5 \right)
\Biggr],
\label{fre}
\eea
where the constants $A_i, B_i,C_i$ and $D_i$ are given in \eqref{coea} -- \eqref{coed}.

Collecting other contributions from ghosts and $Y$, we finally get
\bea
\Gamma^{\rm 1-loop} &=& -\int \frac{d^4 x}{2(4\pi)^2} \Big[
\Big\{ \frac{133}{20} C^2+\Big(10\frac{\la^2}{\xi^2}-5\frac{\la}{\xi}+\frac{5}{36} \Big) \br^2
-\frac{196}{45} E
\nn
&& \hs{10}
+\, \frac{(30\la-\xi)(4\la+\xi)}{12\xi\kappa^2} \br
+ 2\frac{84\la-\xi}{9\kappa^2}\Lambda
+\frac{180\la^2+\xi^2}{72\kappa^4} \Big\} \log \frac{\Lambda_{\rm UV}^2}{\mu^2}
\nn
&& \hs{10}
-\, \Big\{ \frac{144 \la^3-24(7+6\sqrt{\pi}) \la^2\xi +2(59+45\sqrt{\pi})\la\xi^2-(29+14\sqrt{\pi})\xi^3}
{12(3\la-\xi)\xi^2} \br \nn
&& \hs{10}
+\, \frac{\sqrt{\pi}(30\la-\xi)}{12\kappa^2} \Big\} \Lambda_{\rm UV}^2
\nn
&& \hs{10}
-\, \frac{2592\la^4-3456\la^3\xi+3024\la^2\xi^2-1248\la\xi^3+257\xi^4}{72\xi^2(3\la-\xi)^2} \Lambda_{\rm UV}^4
 \Big].
\eea

Note that the coefficient of the Euler term $E$ is independent of any coupling, in particular of its coupling $\rho$.
This is to be expected, because the Euler term is a topological term and is a total derivative itself,
and as such it does not contribute to the Hessian and therefore to quantum effects~\cite{FOP}.
Thus it is a universal result that it is independent of the coupling $\rho$ whatever the approximation is
(beyond one loop).

The correspondence between the cutoff and the dimensional regularization is
 $\log\frac{\Lambda_{\rm UV}^2}{\mu^2} \leftrightarrow \frac{2}{4-d}$ and one can try to compare the results
with the existing literature (see for example, \cite{AB,BuchbinderS}).

Note also that there are strange terms with coefficients $\sqrt{\pi}$ in the quadratic divergences.
Indeed, if we consider the minimal operators $F_1=\Box +P_1$ and $F_2=\Box +P_2$ and consider the quadratic
divergences of $\Tr\log(F_1 F_2)$, it appears that we get such factor of $\sqrt{\pi}$ from the formula $b_2$
in \eqref{qb2}, whereas if we write it as $\Tr\log F_1+\Tr\log F_2$, we do not get $\sqrt{\pi}$ as
is clear from \eqref{2b2} for the same quantity. This clearly indicates that the coefficients of the quadratic
divergences depend on how we calculate. Thus the power divergences are not universal and do not lead to physical
effects like renormalization group scaling. The above results are just those obtained by the naive application
of the formulae but should not be taken seriously.

On the other hand, the logarithmic divergences are universal.
From the coefficients, we can determine the beta functions for the dimensionless couplings:
recall that together with the bare terms, the coefficients of $C^2$ should give the renormalized coupling
\bea
\frac{1}{2 \la_R}=
\frac{1}{2\la_B} -\frac{133}{(4\pi)^2 40} \log\frac{\Lambda_{\rm UV}^2}{\mu^2}.
\eea
Since the bare coupling $\la_B$ does not depend on the renormalization scale $\mu$,
differentiation of the expression with respect to $\log\mu$ gives
\bea
-\frac{1}{2\la_R^2} \mu\frac{d\la_R}{d\mu} = \frac{133}{(4\pi)^2 20}.
\eea
Omitting the subscript $R$, this gives the beta function of the coupling $\la$:
\bea
\mu\frac{d\la}{d\mu} = \b_\la = -\frac{1}{(4\pi)^2}\frac{133}{10}\la^2,
\label{betalambda}
\eea
Similarly we find the beta functions for other dimensionless couplings:
\bea
\beta_\xi &=& -\frac{1}{(4\pi)^2}\left(10\la^2-5\la\xi+\frac{5}{36}\xi^2\right), \nn
\beta_\rho &=& -\frac{1}{(4\pi)^2}\frac{196}{45}\rho^2.
\eea
The first equation~\eqref{betalambda} tells us that the coupling $\la$ goes to zero from positive $\la$.
Similarly the other couplings also go to zero. This is known as asymptotic freedom~\cite{FT1,FT2,AB,BS1,BS2,OP1,OP2}.
However it has recently been discovered that there are fixed point at finite values for these couplings~\cite{FOP}
in addition to these Gaussian fixed points.
See, however, \cite{KO}.

\section{Divergences for $f(R,R_{\mu\nu}^2)$ gravity}

As an interesting case of higher-derivative gravity,
here we present the one-loop divergences for $f(R,R_{\mu\nu})$ gravity on the Einstein space~\cite{OPP}:
\bea
\br_{\mu\nu}=\frac{\br}{d} \bg_{\mu\nu}.
\label{Einstein}
\eea
Moreover we consider a general parametrization of the fluctuations
\bea
\label{gexp}
g_{\mu\nu}=\bg_{\mu\nu}+\delta g_{\mu\nu}\ ,
\eea
where the fluctuation is expanded:
\bea
\label{gammaexp}
\delta g_{\mu\nu}=
\delta g^{(1)}_{\mu\nu}
+\delta g^{(2)}_{\mu\nu}
+\delta g^{(3)}_{\mu\nu}+\ldots \ ,
\eea
where $\delta g^{(n)}_{\mu\nu}$ contains $n$ powers of $h_{\mu\nu}$.
We will parametrize the first two terms of the expansion
as follows:
\bea
\delta g^{(1)}_{\mu\nu}&=&h_{\mu\nu}+m\bg_{\mu\nu}h \ ,
\nonumber\\
\delta g^{(2)}_{\mu\nu}&=&
\omega h_{\mu\rho}h^\rho{}_\nu
+m h h_{\mu\nu}
+m\left(\omega-\frac{1}{2}\right)\bg_{\mu\nu}h^{\alpha\beta}h_{\alpha\beta}
+\frac{1}{2}m^2\bg_{\mu\nu}h^2\ .
\label{deltag}
\eea
It is convenient to use York decomposition
\bea
h_{\mu\nu} = h^{TT}_{\mu\nu} + \bnabla_\mu\xi_\nu + \bnabla_\nu\xi_\mu +
\bnabla_\mu \bnabla_\nu \s -\frac{1}{d} \bg_{\mu\nu} \bnabla^2 \s +
\frac{1}{d} \bg_{\mu\nu} h,
\label{york}
\eea
where
$$
\bnabla^\mu h^{TT}_{\mu\nu} = 0\ ;
\qquad
\bg^{\mu\nu} h^{TT}_{\mu\nu}=0\ ;\qquad
 \bnabla_\mu \xi^\mu=0\ .
$$
We then find that the Hessian is
\bea
S^{(2)}=
\int d^dx\sqrt{\bg}
&& \left[ h^{TT}_{\mu\nu}H^{TT}h^{TT\mu\nu}
+\xi_\mu H^{\xi\xi} \xi^\mu
+\sigma H^{\sigma\sigma} \sigma
+\sigma H^{\sigma h} h
+h H^{h\sigma} \sigma \right.
\nn && \left.
+h H^{hh} h
\right],
\eea
where
\bea
H^{TT} &=& \frac14 \left[ \left\{\bar f_X \left( \lich_2 -\frac{4\br}{d} \right)-\bar f_R\right\}
\left( \lich_2 -\frac{2\br}{d}\right) - (1-2\omega)(1+md) \tilde E \right], \\
H^{\xi\xi} &=& - \frac{(1-2\omega)(1+md)}{2} \left(\lich_1 -\frac{2\br}{d}\right) \tilde E, \\
H^{\sigma\sigma} &=& \frac12 \left(\frac{d-1}{d}\right)^2
\left[
P \lich_0 \left(\lich_0 -\frac{\br}{d-1} \right)
 +Q \lich_0
 \right. \nn
&& \left. \
-\frac{d(1-2\omega)(1+md)}{2(d-1)} \tilde E \right]
\lich_0 \left(\lich_0-\frac{\br}{d-1}\right), \\
H^{\sigma h} &=& \left(\frac{d-1}{d}\right)^2 \frac{1+md}{2} \left[P \left(\lich_0 -\frac{\br}{d-1}\right)
+Q \right]
 \lich_0 \left(\lich_0 -\frac{\br}{d-1}\right), \\
H^{hh} &=& \left(\frac{d-1}{d}\right)^2 \frac{(1+md)^2}{2}
\left[ P \left(\lich_0 - \frac{\br}{d-1} \right)^2
+\, Q\left(\lich_0 -\frac{\br}{d-1}\right)
\right.
\nn
&&
\left.
\qquad\qquad\qquad\qquad\qquad\qquad
+ \frac{d[(1+md)d-2(1-2\omega)]}{4(d-1)^2 (1+md)} \tilde E \right] ,
\eea
where $\lich_2$, $\lich_1$ and $\lich_0$ are the Lichnerowicz Laplacians defined as
\bea
\lich_2 T_{\mu\nu} &=& -\bnabla^2 T_{\mu\nu} +\br_\mu{}^\rho T_{\rho\nu}
+ \br_\nu{}^\rho T_{\mu\rho} -\br_{\mu\rho\nu\s} T^{\rho\s} -\br_{\mu\rho\nu\s} T^{\s\rho}, \nn
\lich_1 V_\mu &=& -\bnabla^2 V_\mu + \br_\mu{}^\rho V_\rho, \nn
\lich_0 S &=& -\bnabla^2 S.
\label{Lichop}
\eea
and the subscripts on $f$ denote derivatives with respect to its arguments:
\bea
f_R=\frac{\partial f}{\partial \br}\ ,\quad
f_X=\frac{\partial f}{\partial X}\ ,\quad
f_{RR}=\frac{\partial^2 f}{\partial \br^2}\ ,\quad
f_{RX}=\frac{\partial^2 f}{\partial \br\partial X}\ ,\quad
f_{XX}=\frac{\partial^2 f}{\partial X^2}\ ,
\eea
with
\bea
X\equiv \br_{\mu\nu}^2,
\eea
We have also used the shorthands
\bea
P&=&\bar f_{RR}+\frac{4}{d^2}\br^2 \bar f_{XX}+4\br \bar f_{RX}
+ \frac{d}{2(d-1)} \bar f_X
\\
Q&=&\frac{d-2}{2(d-1)} \bar f_R + \frac{3d^2-10d+8}{2d(d-1)^2}\br \bar f_X
\eea
and
\bea
\tilde E \equiv \bar f - \frac{2}{d}\br \bar f_R -\frac{4\br^2}{d^2} \bar f_X =0,
\label{onshell2}
\eea
is the field equation evaluated on the Einstein space \eqref{Einstein}.

Our gauge fixing is
\bea
\label{gfaction}
S_{GF}=\frac{1}{2a}\int d^d x \sqrt{\bg}\,\bg^{\mu\nu}F_\mu F_\nu,
\eea
with
\bea
\label{gf}
F_{\mu}=\bar{\bnabla}_{\alpha}{h^{\alpha}}_{\mu}-\frac{\bar b+1}{d}\bar{\bnabla}_{\mu}h\,,
\eea
and $a$ and $\bar b$ are gauge parameters. This can be rewritten as
\bea
\label{gf2}
S_{GF}= - \frac{1}{2a}\int d^dx\sqrt{\bg}
\left[
\xi_\mu\left(\lich_1-\frac{2\br}{d}\right)^2\xi^\mu
+\frac{(d-1-b)^2}{d^2}
\chi\; \lich_0\left(\lich_0-\frac{\br}{d-1-b}\right)^2\chi
\right],\nn
\eea
in terms of the new field
\bea
\chi 
= \frac{(d-1)\lich_0-\br}{(d-1-b)\lich_0-\br}\sigma
+\frac{b(1+dm)}{(d-1-b)\lich_0-\br}h\ ,
\label{chi}
\eea
where $b=\bar b/(1+md)$.

The ghost action contains a nonminimal operator
\bea
S_{gh}=i \int d^dx\sqrt{\bg}\,\bar C^\mu\left(
\delta_\mu^\nu\bnabla^2
+\left(1-2\frac{b+1}{d}\right)\bnabla_\mu\bnabla^\nu+\br_\mu{}^\nu\right)C_\nu ,
\eea
We can rewrite this as
\bea
S_{gh}= i \int d^dx\sqrt{\bg}
\left[ \bar C^{T\mu}\left(\lich_{1}-\frac{2\br}{d}\right)C^T_\mu
+2\frac{d-1-b}{d}
\bar C'^L\left(\lich_{0}-\frac{\br}{d-1-b}\right)C'^L\right].\nn
\eea
in terms of the transverse and longitudinal parts of the ghost field:
\bea
C_\nu=C^T_\nu+\bnabla_\nu C^L
=C^T_\nu+\bnabla_\nu\frac{1}{\sqrt{\lich_{0}}}C'^L ,
\eea
and the same for $\bar C$.
This change of variables has unit Jacobian.
\bea
\label{ghostaction}
S_{gh}= i \int d^dx\sqrt{\bg}
\left[ \bar C^{T\mu}\left(\lich_{1}-\frac{2\br}{d}\right)C^T_\mu
+2\frac{d-1-b}{d}
\bar C'^L\left(\lich_{0}-\frac{\br}{d-1-b}\right)C'^L\right].\nn
\eea

Unless we (1) set $\omega=\frac12$, or (2) put $m=-\frac{1}{d}$ or (3) go on shell,
the effective action is gauge dependent. If we impose the on-shell condition~\eqref{onshell2}.
the effective action is gauge independent and is independent of $\omega$ and $m$.
In this case, we find
\bea
\G &=& \frac12 \Tr\log\left(\lich_2-\frac{4\br}{d} -\frac{\bar f_R}{\bar f_X} \right)
+ \frac12 \Tr\log\left(\lich_2-\frac{2\br}{d} \right)
 \nn
&&
+\frac12 \Tr\log\left(\lich_0-\frac{\br}{d-1}+\frac{Q}{P}\right)
-\frac12 \Tr\log\left(\lich_1-\frac{2\br}{d} \right).
\label{ea}
\eea
If $\bar f_X=0$, the first contribution is absent.

The divergent part of the effective action can be computed by the heat kernel methods.
On an Einstein background in four dimensions, with the help of the heat kernel coefficients for the Lichnerowicz
operator summerized in Appendix~\ref{p=2minimal}, the logarithmically divergent part is found to be~\cite{OPP}
\bea
\Gamma_{\rm log}(\bg) &=& \frac{1}{720(4\pi)^2}
\int d^4x\,\sqrt{\bg}
\log\left(\frac{\Lambda_{\rm UV}^2}{\mu^2}\right)
\left[
-826 \br_{\mu\nu\rho\sigma}^2
+509 \br^2
- \frac{300 \br \bar f_R}{\bar f_X}
- \frac{900 \bar f_R^2}{\bar f_X^2} \right.
\nn
&& \hs{-6}
\left.
+\frac{240\br(3\bar f_R+2\br \bar f_X)}{8\bar f_X+12\bar f_{RR}+48\br \bar f_{RX}+3\br^2 \bar f_{XX}}
-\frac{320(3\bar f_R+2\br \bar f_X)^2}{(8\bar f_X+12\bar f_{RR}+48\br \bar f_{RX}+3\br^2 \bar f_{XX})^2}
\right], \nn
\label{gammaabc}
\eea
where $\Lambda_{\rm UV}$ stands for a cutoff and we introduced a reference mass scale $\mu$.

For the choice
\bea
f(R,X)=\alpha \br^2+\beta X \ ,
\eea
it reduces to
\bea
\Gamma_{\rm log}(\bg) = \frac{1}{(4\pi)^2}
\int d^4x\,\sqrt{\bg}
\log\left(\frac{\Lambda_{\rm UV}^2}{\mu^2}\right)
\left[
-\frac{413}{360} \br_{\mu\nu\rho\sigma}^2
-\frac{1200 \alpha ^2+200 \alpha  \beta -183 \beta ^2}{240 \beta ^2}\br^2 \right] ,\nn
\label{gammaabc1}
\eea
which is the standard universal result in higher-derivative gravity.

On the other hand if we put
\bea
f(R,X)=f(R) \ ,
\eea
we obtain
\bea
\Gamma_{\rm log}(\bg) = \frac{1}{(4\pi)^2}
\int d^4x\,\sqrt{\bg}
\log\left(\frac{\Lambda_{\rm UV}^2}{\mu^2}\right)
\left[
-\frac{71}{120} \br_{\mu\nu\rho\sigma}^2
+\frac{433}{1440}\br^2
+\frac{\bar f_R \br}{12\bar f_{RR}}
-\frac{\bar f_R^2}{36\bar f_{RR}^2}\right],\nn
\label{gammaabc2}
\eea
which agrees with the results of \cite{RS,Ohta}.

\section*{Acknowledgments}

We would like to thank Andrei Barvinsky, Roberto Percacci and Ilya Shapiro for valuable discussions.
This work was supported in part by the Grant-in-Aid for Scientific Research Fund of the JSPS (C) No. 16K05331,
No. 20K03980, and by the Ministry of Science and Technology, R. O. C. (Taiwan) under the grant MOST 111-2811-M-008-024.

\renewcommand{\thesection}{\Alph{section}}
\setcounter{section}{1}

\section*{Appendix}

\subsection{Expansion of curvatures up to second order}
\label{expansion}

Here we summarize our conventions and formulae necessary in the text.
We give these such that they are valid for any dimension $d$.

Our signature of the metric is $(-,+,\dots +)$ and the curvature tensors are given as
\bea
R^\a{}_{\b\mu\nu} &=&
\pa_\mu \G^{\a}_{\b\nu} - \pa_\nu \G^{\a}_{\b\mu}
+ \G^{\a}_{\mu\la} \G^{\la}_{\b\nu} - \G^{\a}_{\nu\la} \G^{\la}_{\b\mu}, \nn
R_{\mu\nu} &=& \br^\a{}_{\mu\a\nu}.
\eea
The backgrounds are denoted with overbar.
Expansion around the background gives
\bea
\G^\a_{\mu\nu}
&=& \bar \G^\a_{\mu\nu} + \G^{\a (1)}_{\mu\nu} + \G^{\a(2)}_{\mu\nu},
\eea
where
\bea
\G^{\a (1)}_{\mu\nu} &=& \frac12 (\bnabla_\nu h^\a{}_\mu
 +\bnabla_\mu h^\a{}_\nu-\bnabla^\a h_{\mu\nu}), \\
\G^{\a(2)}_{\mu\nu} &=& -\frac12 h^{\a\b} (\bnabla_\nu h_{\mu\b}
+\bnabla_\mu h_{\nu\b}-\bnabla_\b h_{\mu\nu}).
\eea
Note that
\bea
\sqrt{-g} = \sqrt{-\bg} \Big[ 1+\frac{1}{2}h
+ \frac{1}{8}(h^2-2 h_{\mu\nu}^2) + O(h^3) \Big],
\eea
where $h\equiv h_\mu{}^\mu$.
We find, to the second order,
\bea
R^\mu{}_{\nu\a\b} &=& \br^\mu{}_{\nu\a\b} + \br^\mu{}_{\nu\a\b}^{(1)}
+ \br^\mu{}_{\nu\a\b}^{(2)}, \nn
R^\mu{}_{\nu\a\b}^{(1)} &=& \frac12 (\bnabla_\a\bnabla_\nu h^\mu_\b
- \bnabla_\a\bnabla^\mu h_{\nu\b} - \bnabla_\b \bnabla_\nu h^\mu_\a
+ \bnabla_\b\bnabla^\mu h_{\nu\a})
\nn && +\frac12 \br_{\nu\c\a\b} h^{\mu\c}
+\frac12 \br^\mu{}_{\c\a\b} h^\c_\nu,~~~ \\
R^\mu{}_{\nu\a\b}^{(2)}
&=& -\frac12 h^{\mu\c}\bnabla_\a (\bnabla_\b h_{\nu\c}+\bnabla_\nu h_{\b\c}
-\bnabla_\c h_{\nu\b})
-\frac14 \bnabla_\a h^{\mu\c}(\bnabla_\b h_{\nu\c}+\bnabla_\nu h_{\b\c}
-\bnabla_\c h_{\nu\b}) \nn
&&\hs{-2} +\; \frac14 \bnabla_\c h^\mu_\a (\bnabla_\b h^\c_\nu+\bnabla_\nu h^\c_\b
-\bnabla^\c h_{\nu\b})
-\frac14 \bnabla^\mu h_{\a\c} (\bnabla_\b h^\c_\nu+\bnabla_\nu h^\c_\b
-\bnabla^\c h_{\nu\b}) \nn
&& - (\a \leftrightarrow \b).
\eea
Similarly
\bea
R^{(1)}_{\mu\nu} &=& -\frac12 (\bnabla_\mu \bnabla_\nu h
- \bnabla_\mu h_{\nu} - \bnabla_\nu h_\mu + \Box h_{\mu\nu})
- \br_{\a\mu\b\nu}h^{\a\b}+\frac12 \br_{\mu\a}h^\a_\nu +\frac12 \br_{\nu\a} h^\a_\mu
, \nn
R^{(2)}_{\mu\nu} &=& \frac12 \bnabla_\mu(h^{\a\b} \bnabla_\nu h_{\a\b})
-\frac12 \bnabla_\a \{h^{\a\b}( \bnabla_\mu h_{\nu\b}+ \bnabla_\nu h_{\mu\b}
- \bnabla_\b h_{\mu\nu}) \} \nn
&& \hs{-10}
- \frac14 (\bnabla_\mu h^\b_\a+ \bnabla_\a h^\b_\mu -\bnabla^\b h_{\a\mu})
(\bnabla_\b h^\a_\nu + \bnabla_\nu h^\a_\b -\bnabla^\a h_{\b\nu})
+ \frac14 \bnabla_\a h (\bnabla_\mu h^\a_\nu + \bnabla_\nu h_\mu^\a -\bnabla^\a h_{\mu\nu}), \nn
R^{(1)} &=& \bnabla_\mu h^\mu -\Box h - \br_{\mu\nu} h^{\mu\nu}, \nn
R^{(2)} &=& \frac12 \bnabla_\mu(h^{\a\b} \bnabla^\mu h_{\a\b})
-\frac12 \bnabla_\a \{h^{\a\b}( 2 h_\b - \bnabla_\b h) \}
- \frac14 (\bnabla_\mu h^\b_\a+ \bnabla_\a h^\b_\mu -\bnabla^\b h_{\a\mu}) \bnabla_\b h^{\a\mu}
 \nn
&& \hs{-10}
+\; \frac14 ( 2 h^\a -\bnabla^\a h)\bnabla_\a h
+\frac12 h^{\a\b}\bnabla_\a \bnabla_\b h
-\frac12 h_\a^\mu \bnabla_\b ( \bnabla^\a h_\mu^\b +\bnabla_\mu h^{\a\b}
-\bnabla^\b h^\a_\mu) + \br_{\mu\nu} h^\mu_\a h^{\nu\a} \nn
&=&
\frac34 \bnabla_\a h_{\mu\nu} \bnabla^\a h^{\mu\nu} +h_{\mu\nu} \Box h^{\mu\nu}
-h_\mu^2 + h_\mu \bnabla^\mu h -2 h_{\mu\nu} \bnabla^\mu h^\nu
+h_{\mu\nu} \bnabla^\mu \bnabla^\nu h \nn
&&
-\; \frac12 \bnabla_\mu h_{\nu\a} \bnabla^\a h^{\mu\nu} -\frac14 \bnabla_\mu h \bnabla^\mu h
+\br_{\a\b\c\d} h^{\a\c} h^{\b\d},
\eea
where $h_\mu\equiv \bnabla^\nu h_{\mu\nu}$.
Note that $\bg^{\mu\nu} R^{(1)}_{\mu\nu} \neq R^{(1)}$,
because the latter has additional contribution from $h^{\mu\nu} \br_{\mu\nu}$.
When total derivative terms are dropped, $R^{(2)}$ makes the contribution to the action
\bea
R^{(2)} \simeq
\frac14 ( h_{\mu\nu}\Box h^{\mu\nu} +h \Box h + 2 h_\mu^2
+2 \br_{\a\b}h^{\a\c} h^\b_\c + 2 \br_{\a\b\c\d}h^{\a\c} h^{\b\d}).
\eea
We use the notation $\simeq$ to denote equality up to total derivatives.

\newpage
\subsection{Heat kernel coefficients for $p=2$ minimal operator}
\label{p=2minimal}

For minimal operator $\Delta=-\bnabla^2+\bm E$, the general formulae for any spin are
\bea
&& b_0 = \tr\, \hat 1, \qquad
b_2 = \frac{1}{6}\br\, \tr\, \hat 1 - \tr\, \bm E, \nn
&& b_4 = \frac{1}{180} \Big( \br_{\mu\nu\a\b}^2 - \br_{\mu\nu}^2+\frac{5}{2}\br^2+6\bnabla^2 \br \Big) \tr\,\hat 1
+\frac{1}{2} \tr\, \bm E^2
\nn && \hs{15}
-\frac{1}{6}\br\, \tr\, \bm E +\frac{1}{12} \tr\, \Omega_{\mu\nu}^2
-\frac{1}{6} \bnabla^2 \tr\, \bm E,
\label{master}
\eea
where $\Omega_{\mu\nu}=[\bnabla_\mu,\bnabla_\nu]$ is the curvature from the covariant derivatives for each spin,
and the traces should be taken using the identity $\eta_{\mu\nu}$ for the vector and \eqref{id} for
the symmetric tensor. The formulae in subsection~\ref{p=2nonminimalvector} are for spin 1, but with an additional
nonminimal term.

Using these formulae~\eqref{master} to the Lichnerowicz Laplacians~\eqref{Lichop}, we find,
for spin 0
\bea
&& b_0(\lich_0)=1, \qquad
b_2(\lich_0)=\frac16 \br, \nn
&&
b_4(\lich_0)=\frac{1}{180} \Big( \br_{\mu\nu\a\b}^2 - \br_{\mu\nu}^2+\frac{5}{2}\br^2+6\bnabla^2 \br \Big),
\label{unb0}
\eea
and for spin 1
\bea
&& b_0(\lich_1) = d, \qquad
b_2(\lich_1) = \frac{d-6}{6}\br,  \nn
&&
b_4 = \frac{d-15}{180} \br_{\mu\nu\a\b}^2 - \frac{d-90}{180} \br_{\mu\nu}^2+\frac{d-12}{72} \br^2
+\frac{d-5}{30}\bnabla^2 \br,
\label{unb1}
\eea
since $\Omega_{\mu\nu}A_\a=\br_{\mu\nu\a\b}A^\b\equiv (\Omega_{\mu\nu})_{\a\b}A^\b$ and
\bea
\tr\, \Omega_{\mu\nu}^2 = \br_{\mu\nu\a\rho} \br^{\mu\nu\rho\a}=-\br_{\mu\nu\a\b}^2
\eea
For spin 2, we find
\bea
b_0(\lich_2) &=& \frac{d(d+1)}{2}, \qquad
b_2(\lich_2) = \frac{d^2-11d-24}{12}\br,  \nn
b_4(\lich_2) &=& \frac{d^2-29d+480}{360} \br_{\mu\nu\a\b}^2 - \frac{d^2-359d-1080}{360} \br_{\mu\nu}^2 \nn
&&
+\frac{d^2-23d-48}{144} \br^2+\frac{d^2-9d-20}{60} \bnabla^2 \br,
\label{unb2}
\eea
since
$(\Omega_{\mu\nu}^2)_{\a\b,\rho\s}= \br_{\mu\nu\a\c} \br^{\mu\nu\c}{}_\rho g_{\b\s}
+ \br_{\mu\nu\a\rho} \br^{\mu\nu}{}_{\b\s}
+ \br_{\mu\nu\b\s} \br^{\mu\nu}{}_{\a\rho}
+ \br_{\mu\nu\b\c} \br^{\mu\nu\c}{}_{\s} g_{\a\rho}$ and
\bea
\tr(\Omega^2_{\mu\nu}) &=& \frac{1}{2}\Big(\{\mbox{sum over }\a=\rho,\b=\s \}+\{\mbox{sum over }\a=\s,\b=\rho \}\Big)
\nn
&=& -(d+2)\br_{\mu\nu\a\c}^2.
\eea
These are the results when the fields do not have any constraints. If the fields have constraint such as transverse,
suitable subtraction is needed for spins 1 and 2.

Spin 0 does not have any constraint, so their heat kernel coefficients are the same as above.
For the transverse spin 1, we have $b_n(\lich_1^T)=b_n(\lich_1)-b_n(\lich_0)$. Hence
\bea
&& b_0(\lich_1^T)= d-1,\quad
b_2(\lich_1^T)=\frac{d-7}{6}\br,\quad \nn
&& b_4(\lich_1^T)= \frac{d-16}{180} \br_{\mu\nu\rho\s}^2 -\frac{d-91}{180} \br_{\mu\nu}^2
+\frac{d-13}{72}\br^2+\frac{d-6}{30}\bnabla^2 \br.
\label{lich1}
\eea
For the Einstein space, $b_4(\lich_1^T)$ reduces to
\bea
b_4(\lich_1^T)= \frac{d-16}{180} \br_{\mu\nu\rho\s}^2 +\frac{5d^2-67d+182}{360 d} \br^2+\frac{d-6}{30}\bnabla^2 \br.
\eea

Similarly those for transverse and traceless spin 2, we have
$b_n(\lich_2^{TT})=b_n(\lich_2)-b_n(\lich_1^T)-2b_n(\lich_0)$, so
\bea
b_0(\lich_2^{TT}) &=& \frac{(d+1)(d-2)}{2},\quad
b_2(\lich_2^{TT}) = \frac{(d+1)(d-14)}{12}\br,\quad \nn
b_4(\lich_2^{TT}) &=& \frac{d^2-31d+508}{360} \br_{\mu\nu\rho\s}^2 -\frac{d^2-361d-902}{360} \br_{\mu\nu}^2 \nn
&& +\frac{(d+1)(d-26)}{144}\br^2+\frac{(d+1)(d-12)}{60}\bnabla^2 \br.
\label{lich2}
\eea
For the Einstein space, $b_4(\lich_2^{TT})$ reduces to
\bea
b_4(\lich_2^{TT})&=& \frac{d^2-31d+508}{360} \br_{\mu\nu\rho\s}^2 +\frac{5d^3-127d^2+592d+1804}{720 d} \br^2
\nn &&
+\frac{(d+1)(d-12)}{60}\bnabla^2 \br.
\eea

Finally we also need the following formulae which also follow from \eqref{master}:
\bea
b_0(\Delta+aR) &=& b_0(\Delta), \nn
\label{shift}
b_2(\Delta+aR) &=& b_2(\Delta) -aR b_0(\Delta) , \\
b_4(\Delta+aR) &=& b_4(\Delta) -aR b_2(\Delta)+\frac12 a^2 R^2 b_0(\Delta) . \nonumber
\eea

We can evaluate the one-loop divergent part~\eqref{div2} of the effective action~\eqref{ea}
by using eqs.~\eqref{unb0}, \eqref{lich1} -- \eqref{shift}.


\newpage
\begin{thebibliography}{99}

\bibitem{TV}
G.~'t Hooft and M.~J.~G.~Veltman,
``One loop divergencies in the theory of gravitation,''
Ann. Inst. H. Poincare Phys. Theor. A \textbf{20}, 69 (1974)

\bibitem{GS}
M.~H.~Goroff and A.~Sagnotti,
``The Ultraviolet Behavior of Einstein Gravity,''
Nucl. Phys. B \textbf{266}, 709 (1986)

\bibitem{Stelle}
K.~S.~Stelle,
``Renormalization of Higher Derivative Quantum Gravity,''
Phys. Rev. D \textbf{16}, 953 (1977)

\bibitem{D}
J.~F.~Donoghue and G.~Menezes,
``Unitarity, stability and loops of unstable ghosts,''
Phys. Rev. D \textbf{100} (2019), 105006 [arXiv:1908.02416 [hep-th]].

\bibitem{F}
R.~P.~Feynman,
``Quantum theory of gravitation,''
Acta Phys. Polon. \textbf{24}, 697 (1963)

\bibitem{DeWitt}
B.~S.~DeWitt,
``Quantum Theory of Gravity. 1. The Canonical Theory,''
Phys. Rev. \textbf{160}, 1113 (1967);
``Quantum Theory of Gravity. 2. The Manifestly Covariant Theory,''
Phys. Rev. \textbf{162}, 1195 (1967);
``Quantum Theory of Gravity. 3. Applications of the Covariant Theory,''
Phys. Rev. \textbf{162}, 1239 (1967)

\bibitem{FP}
L.~D.~Faddeev and V.~N.~Popov,
``Feynman Diagrams for the Yang-Mills Field,''
Phys. Lett. B \textbf{25}, 29 (1967)

\bibitem{JT}
J.~Julve and M.~Tonin,
``Quantum Gravity with Higher Derivative Terms,''
Nuovo Cim. B \textbf{46}, 137 (1978)

\bibitem{FT1}
E.~S.~Fradkin and A.~A.~Tseytlin,
``Renormalizable Asymptotically Free Quantum Theory of Gravity,''
Phys. Lett. B \textbf{104}, 377 (1981)

\bibitem{FT2}
E.~S.~Fradkin and A.~A.~Tseytlin,
``Renormalizable asymptotically free quantum theory of gravity,''
Nucl. Phys. B \textbf{201}, 469 (1982)

\bibitem{BC} N.~H.~Barth and S.~M.~Christensen,
``Quantizing Fourth Order Gravity Theories. 1. The Functional Integral,''
Phys. Rev. D \textbf{28}, 1876 (1983)

\bibitem{Ohta:2020}
N.~Ohta,
``General Procedure of Gauge Fixings and Ghosts,''
Phys. Lett. B \textbf{811}, 135965 (2020)
[arXiv:2010.11314 [hep-th]].

\bibitem{G}
V.~P.~Gusynin,
``Seeley-Gilkey Coefficients for the Fourth Order Operators on a Riemannian Manifold,''
Nucl. Phys. B \textbf{333}, 296 (1990)

\bibitem{GK}
V.~P.~Gusynin and V.~V.~Kornyak,
``Complete computation of DeWitt-Seeley-Gilkey coefficient E(4) for nonminimal operator on curved manifolds,''
Fund. Appl. Math. \textbf{5}, 649(1999)
[arXiv:math/9909145 [math]].

\bibitem{BV}
A.~O.~Barvinsky and G.~A.~Vilkovisky,
``The Generalized Schwinger-Dewitt Technique in Gauge Theories and Quantum Gravity,''
Phys. Rept. \textbf{119}, 1 (1985)

\bibitem{AB}
I.~G.~Avramidi and A.~O.~Barvinsky,
``Asymptotic freedom in higher derivative quantum gravity,''
Phys. Lett. B \textbf{159}, 269 (1985)

\bibitem{BS1}
G.~de Berredo-Peixoto and I.~L.~Shapiro,
``Conformal quantum gravity with the Gauss-Bonnet term,''
Phys. Rev. D \textbf{70}, 044024 (2004) [arXiv:hep-th/0307030 [hep-th]].

\bibitem{BS2}
G.~de Berredo-Peixoto and I.~L.~Shapiro,
``Higher derivative quantum gravity with Gauss-Bonnet term,''
Phys. Rev. D \textbf{71}, 064005 (2005) [arXiv:hep-th/0412249 [hep-th]].

\bibitem{OP1}
N.~Ohta and R.~Percacci,
``Higher Derivative Gravity and Asymptotic Safety in Diverse Dimensions,''
Class. Quant. Grav. \textbf{31}, 015024 (2014)
[arXiv:1308.3398 [hep-th]].

\bibitem{OP2}
N.~Ohta and R.~Percacci,
``Ultraviolet Fixed Points in Conformal Gravity and General Quadratic Theories,''
Class. Quant. Grav. \textbf{33}, 035001 (2016) [arXiv:1506.05526 [hep-th]].

\bibitem{FOP}
K.~Falls, N.~Ohta and R.~Percacci,
``Towards the determination of the dimension of the critical surface in asymptotically safe gravity,''
Phys. Lett. B \textbf{810}, 135773 (2020) [arXiv:2004.04126 [hep-th]].

\bibitem{BuchbinderS}
I.~L.~Buchbinder and I.~Shapiro,
``Introduction to Quantum Field Theory with Applications to Quantum Gravity,'' Oxford University Press,
Oxford, 2021.

\bibitem{KO}
H.~Kawai and N.~Ohta,
``Wave function renormalization and flow of couplings in asymptotically safe quantum gravity,''
Phys. Rev. D \textbf{107}, 126025 (2023) [arXiv:2305.10591 [hep-th]].

\bibitem{OPP}
N.~Ohta, R.~Percacci and A.~D.~Pereira,
``$f(R, R_{\mu\nu}^2)$ at one loop,''
Phys. Rev. D \textbf{97}, 104039 (2018) [arXiv:1804.01608 [hep-th]].

\bibitem{RS}
M.~S.~Ruf and C.~F.~Steinwachs,
``One-loop divergences for $f(R)$ gravity,''
Phys. Rev. D \textbf{97}, 044049 (2018) [arXiv:1711.04785 [gr-qc]].

\bibitem{Ohta}
N.~Ohta, ``Quantum Equivalence of $f(R)$ Gravity and Scalar-tensor Theories in the Jordan and Einstein Frames,''
PTEP {\bf 2018}  (2018) no.3,  033B02 [arXiv:1712.05175 [hep-th]].

\end{thebibliography}
\end{document}